\documentclass[cameraready]{Interspeech}

\title{HumDial-EIBench: A Human-Recorded Multi-Turn Emotional Intelligence Benchmark for Audio Language Models}

\author[affiliation={1}, equalcontribution]{Shuiyuan}{Wang}
\author[affiliation={1}, equalcontribution]{Zhixian}{Zhao}
\author[affiliation={1}]{Hongfei}{Xue}
\author[affiliation={1}]{Chengyou}{Wang}
\author[affiliation={2}]{Shuai}{Wang}
\author[affiliation={3}]{Hui}{Bu}
\author[affiliation={3}]{Xin}{Xu}
\author[affiliation={1}, correspondingauthor]{Lei}{Xie}

\address{
    $^1$ Audio, Speech and Language Processing Group (ASLP@NPU), School of Computer Science, Northwestern Polytechnical University, Xi’an, China \\
    $^2$ Nanjing University, China \\
    $^3$ AISHELL, China
}

\email{wangshuiyuan@mail.nwpu.edu.cn, zxzhao@mail.nwpu.edu.cn}


\usepackage{comment}
\usepackage{multirow}

\begin{document}

\maketitle

\begin{abstract}
\vspace{-0.2cm}
Evaluating the emotional intelligence (EI) of audio language models (ALMs) is critical. However, existing benchmarks mostly rely on synthesized speech, are limited to single-turn interactions, and depend heavily on open-ended scoring. This paper proposes HumDial-EIBench, a comprehensive benchmark for evaluating ALMs' EI. Using real-recorded human dialogues from the ICASSP 2026 HumDial Challenge, it reformulates emotional tracking and causal reasoning into multiple-choice questions with adversarial distractors, mitigating subjective scoring bias for cognitive tasks. It retains the generation of empathetic responses and introduces an acoustic-semantic conflict task to assess robustness against contradictory multimodal signals. Evaluations of eight ALMs reveal that most models struggle with multi-turn emotional tracking and implicit causal reasoning. Furthermore, all models exhibit decoupled textual and acoustic empathy, alongside a severe text-dominance bias during cross-modal conflicts.
\end{abstract}

\vspace{-0.2cm}
\section{Introduction}

Traditional spoken dialogue systems rely on cascaded architectures (automatic speech recognition $\to$ large language model $\to$ text-to-speech), where the intermediate text transcription inevitably discards critical paralinguistic cues, such as intonation and emotion. The recent paradigm shift toward end-to-end audio language models (ALMs)~\cite{gpt4o, gemini2.5, qwen-audio,qwen2audio, speechgpt} enables models to process continuous audio signals directly, theoretically maintaining a unified representation of semantic content and acoustic paralinguistic features to achieve native emotional intelligence (EI)~\cite{e-chat, osum, osum-echat, emoomni}. However, a critical question remains: \textit{do these advanced models genuinely perceive acoustic emotion, or do they still primarily process audio as a proxy for text transcription?}

Despite the rapid development of ALMs, existing emotional evaluation benchmarks exhibit limitations that obscure genuine EI capabilities. First, most multi-turn dialogue benchmarks rely entirely on TTS-synthesized speech. This creates a pseudo-multi-turn effect that disrupts natural, continuous emotional evolution and lacks authentic acoustic nuances. Second, although some benchmarks incorporate authentic human recordings, they strictly confine evaluation to static, single-turn emotion recognition, failing to evaluate emotional trajectory tracking and implicit causal reasoning across multiple conversational turns. Third, current frameworks rely heavily on unstable LLM-as-a-judge scoring, which introduces severe subjective instability and fails to objectively distinguish cognitive reasoning errors from superficial generation fluency. Furthermore, existing evaluations consistently omit the systematic examination of complex cross-modal contradictions, such as acoustic-semantic conflicts~\cite{case}. These methodological gaps mask fundamental weaknesses in multimodal alignment, preventing the accurate assessment of true emotional intelligence. Table~\ref{tab:benchmark_comparison} summarizes this comparison.

To overcome these evaluation bottlenecks, this paper proposes HumDial-EIBench\footnote{ \urlstyle{same} \url{https://github.com/ASLP-lab/HumDial-EIBench}}, a comprehensive benchmark for systematically diagnosing the emotional intelligence of ALMs. Based on the authentic human-recorded multi-turn dialogues from the ICASSP 2026 HumDial Challenge~\cite{icassp}, this benchmark reformulates traditional open-ended emotional understanding tasks into objective multiple-choice questions (MCQs) utilizing adversarial distractors. This design bypasses superficial text generation fluency to accurately measure underlying contextual reasoning capabilities. Additionally, alongside the empathetic response generation task, HumDial-EIBench introduces a dedicated task for acoustic-semantic conflict recognition. This addition systematically evaluates cross-modal integration and perceptual robustness when text semantics contradict acoustic emotions. The main contributions are threefold:
\begin{itemize}
    \item HumDial-EIBench provides a comprehensive evaluation framework grounded in real human multi-turn dialogues, covering emotional trajectory detection, causal reasoning, empathetic response generation, and acoustic-semantic conflict recognition.
    \item The adversarial MCQ formulation bypasses the subjective instability of LLM judges for contextual comprehension, effectively isolating cognitive reasoning deficits from generative expression flaws.
    \item Systematic evaluations of eight mainstream ALMs identify critical performance gaps: current models struggle with multi-turn temporal modeling and implicit causal reasoning, and they exhibit a clear divergence between textual and acoustic empathy alongside a severe text-dominance bias when resolving cross-modal conflicts.
\end{itemize}

\begin{table*}[t]
  \caption{Comparison of HumDial-EIBench with existing benchmarks. A-S Conflict denotes Acoustic-Semantic Conflict.}
  \label{tab:benchmark_comparison}
  \centering
  \footnotesize
  \begin{tabular}{lccccccc}
    \toprule
    \multirow{3}{*}{\textbf{Benchmark}} & \multirow{3}{*}{\textbf{\begin{tabular}[c]{@{}c@{}}Input\\Source\end{tabular}}} & \multirow{3}{*}{\textbf{Turns}} & \multicolumn{4}{c}{\textbf{Targeted Capabilities}} & \multirow{3}{*}{\textbf{\begin{tabular}[c]{@{}c@{}}Eval\\Method\end{tabular}}} \\
    \cmidrule(lr){4-7}
    & & & \textbf{Emotion} & \textbf{Causal} & \textbf{Empathetic} & \textbf{A-S} & \\
    & & & \textbf{Trajectory} & \textbf{Reasoning} & \textbf{Response} & \textbf{Conflict} & \\
    \midrule
    Dynamic-SUPERB~\cite{dynamic-superb} & Mixed & Single & $\times$ & $\times$ & $\times$ & $\times$ & Obj.  \\
    AIR-Bench~\cite{air-bench}      & Human-rec.   & Single & $\times$ & $\times$ & $\times$ & $\times$ & LLM / Human \\
    HPSU~\cite{hpsu}           & Human-rec. & Single & $\times$ & $\times$ & $\times$ & $\times$ & Obj. \\
    VoiceBench~\cite{voicebench}   & Mixed & Single & $\times$ & $\times$ & $\times$ & $\times$ & Obj. / LLM \\
    URO-Bench~\cite{uro-bench}       & Mixed   & Multi & $\times$ & $\times$ & $\times$ & $\times$ & LLM \\
    ParaS2S~\cite{paras2s}      & Mixed & Single & $\times$ & $\times$ & $\times$ & \checkmark & LLM \\
    VoxDialogue~\cite{voxdialogue} & TTS & Multi & $\times$ & $\times$ & \checkmark & $\times$ & LLM \\
    MTalk-Bench~\cite{mtalk-bench}     & Human-rec. & Multi & $\times$ & \checkmark & \checkmark & $\times$ & LLM / Human \\
    Multi-Bench~\cite{multi-bench}     & Mixed & Multi & \checkmark & $\times$ & \checkmark & $\times$ & LLM \\
    \midrule
    \textbf{HumDial-EIBench (Ours)} & \textbf{Human-rec.} & \textbf{Multi} & \textbf{\checkmark} & \textbf{\checkmark} & \textbf{\checkmark} & \textbf{\checkmark} & \textbf{Obj. / LLM / Human} \\
    \bottomrule
  \end{tabular}
  \vspace{-0.3cm}
\end{table*}

\vspace{-0.2cm}
\section{Related Work}

\textbf{Audio Language Models.} Traditional spoken dialogue systems adopt a cascaded ASR$\to$LLM$\to$TTS architecture, where the intermediate text transcription inevitably discards paralinguistic cues such as intonation and emotion. End-to-end ALMs---including open-source models like Moshi~\cite{moshi}, Qwen2.5-Omni~\cite{Qwen2.5-Omni}, and Qwen3-Omni~\cite{Qwen3-Omni}, alongside closed-source architectures such as GPT-4o~\cite{gpt4o}---have emerged to address this limitation. By directly processing continuous audio signals, these models maintain unified representations of both semantic content and acoustic paralinguistic features, establishing the technical foundation for native emotional intelligence tasks.

\noindent\textbf{Evaluation Benchmarks.} The development of ALMs necessitates corresponding evaluation frameworks. Early benchmarks, such as AIR-Bench and Dynamic-SUPERB~\cite{air-bench, dynamic-superb, voicebench}, primarily test ASR precision and instruction-following. Subsequent benchmarks, including URO-Bench, ParaS2S, and VoxDialogue~\cite{uro-bench, paras2s, voxdialogue}, introduce complex conversational reasoning but rely entirely on TTS-synthesized speech. While HPSU and ISA-Bench~\cite{hpsu, isa-bench} incorporate authentic human recordings, they restrict evaluation to static, single-turn emotion recognition, precluding the assessment of cross-turn emotional dynamics. For multi-turn emotional interactions, Multi-Bench and MTalk-Bench~\cite{multi-bench, mtalk-bench} assess emotional trajectory tracking and empathy. However, these benchmarks exhibit two fundamental limitations: their reliance on TTS synthesis generates a \textit{pseudo-multi-turn} effect that disrupts natural continuous emotional evolution, and their widespread use of LLM-as-a-Judge scoring introduces severe subjective instability. Furthermore, although specific studies~\cite{case, c2ser} explore acoustic-semantic mismatches within traditional speech emotion recognition (SER), comprehensive ALM benchmarks omit systematic cross-modal contradiction evaluation. HumDial-EIBench resolves these structural gaps by assessing ALM emotional intelligence through authentic human multi-turn dialogues, objective multiple-choice formulations, and an explicit acoustic-semantic conflict task.

\vspace{-0.2cm}
\section{HumDial-EIBench}

\begin{figure*}[t]
  \centering
  \includegraphics[width=0.95\textwidth]{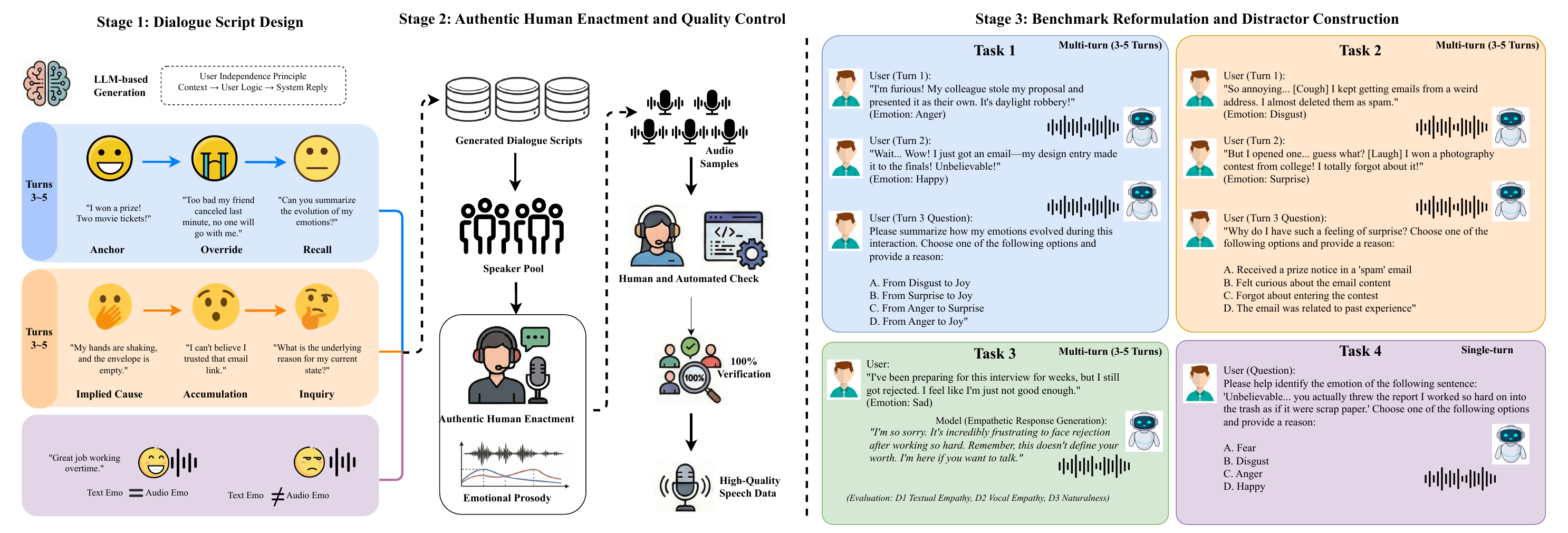}
    \vspace{-0.2cm}
  \caption{Data construction and task overview of HumDial-EIBench. \textbf{Left:} Three-stage pipeline---\textbf{Stage~1}: Dialogue Script Design; \textbf{Stage~2}: Authentic Human Enactment and Quality Control; \textbf{Stage~3}: Multiple-Choice Reformulation and Distractor Construction. \textbf{Right:} Representative examples of the four tasks.}
  \label{fig:humdial_pipeline}
\end{figure*}

HumDial-EIBench is directly built upon the test set of the ICASSP 2026 HumDial Challenge. To ensure both controllability and realism, the foundational data were created by designing specific dialogue scenarios and speaker turns, which were then naturalistically enacted. This paper extends this foundation by reformulating the open-ended challenges into an objective evaluation framework and introducing a cross-modal conflict task. Figure~\ref{fig:humdial_pipeline} details the complete three-stage data construction pipeline and presents representative examples for the four targeted evaluation tasks.

\vspace{-0.2cm}
\subsection{Task formulation and statistics}
The proposed benchmark comprises four tasks constructed directly from high-quality human recordings. Table~\ref{tab:data_stats} summarizes the data scale and designated evaluation metrics. While Tasks~1, 2, and 3 use multi-turn conversational audio, primarily sourced from the HumDial Challenge test set, this benchmark reformulates the original open-ended trajectory and reasoning tasks (Tasks~1 and 2) into objective multiple-choice questions (MCQs). Task~3 retains its open-ended generative format. Furthermore, Task~4 is newly collected and designed using single-turn samples to evaluate acoustic-semantic consistency and contradiction. All tasks include Chinese (CN) and English (EN) subsets, yielding a total of 1{,}077 evaluation samples.
\begin{itemize}
    \item \textbf{Task~1 (Emotional Trajectory Detection)} examines emotional memory in dialogues, requiring the ALM to select the correct trajectory of user emotional shifts over time (e.g., $E_{t1} \to E_{t2} \to E_{t3}$) from the provided options.
    \item \textbf{Task~2 (Implicit Causal Reasoning)} requires inferring the unstated root cause of the current emotion from scattered context. Formulating this task as an MCQ minimizes scoring subjectivity while strictly evaluating deep contextual understanding.
    \item \textbf{Task~3 (Empathetic Response Generation)} assesses both the semantic depth (text empathy) and the acoustic appropriateness (acoustic empathy) of the generated response.
    \item \textbf{Task~4 (Acoustic-Semantic Conflict)} targets scenarios where textual semantics contradict acoustic emotions (e.g., sarcasm). It tests whether the model can identify the true emotional state primarily from acoustic cues despite textual interference, evaluating its robustness against text-dominance bias.
\end{itemize}

\begin{table}[t]
  \caption{Data and evaluation overview of HumDial-EIBench. Gen.: Generation.}
  \label{tab:data_stats}
  \centering
  \footnotesize
  \setlength{\tabcolsep}{3pt}
  \begin{tabular}{llcccc}
    \toprule
    \textbf{Task} & \textbf{Type} & \textbf{Turns} & \textbf{Dialogue (CN/EN)} & \textbf{Utterance} & \textbf{Metric} \\
    \midrule
    Task 1   & MCQ  & 3--5 & 150/150 & 1200 & ACC \\
    Task 2    & MCQ  & 3--5 & 134/149 & 1132 & ACC \\
    Task 3   & Gen. & 3--5 & 144/150 & 1170 & LLM/Hum \\
    Task 4 & MCQ  & 1     & 100/100 & 200 & ACC \\
    \bottomrule
  \end{tabular}
  \vspace{-0.2cm}
\end{table}

\vspace{-0.2cm}
\subsection{Data construction pipeline}
The data construction pipeline consists of three sequential stages.

\vspace{-0.2cm}
\subsubsection{Dialogue script design}
In the foundational dataset, dialogue scenarios and multi-turn speaker scripts were pre-generated using an auxiliary LLM based on specified scenario instructions. To ensure rigorous evaluation and acoustic consistency, these scripts were fixed offline rather than generated dynamically by the evaluated models. The scripts fall into three principal categories:

\begin{itemize}
    \item \textbf{Trajectory-Oriented (Tasks~1 \& 3):} These scripts establish an initial emotion (e.g., ``slightly sad''), introduce a mid-dialogue turning point (improvement or deterioration), and conclude with a neutral inquiry to form a traceable emotional evolution (e.g., $E_{t1} \to E_{t2} \to E_{t3}$ in a three-turn example).
    \item \textbf{Reasoning-Oriented (Tasks~2 \& 3):} The user disperses implicit emotional trigger details across multiple turns (e.g., chronic overtime or family conflicts) without explicit affect labels. The task forces the model to infer the root cause from the accumulated historical context.
    \item \textbf{Conflict-Oriented (Task~4):} Single-sentence text scripts are designed for human speakers to deliver in two distinct manners: one where vocal emotion matches the text sentiment (consistent), and another where vocal emotion contradicts the text sentiment (conflict), generating paired contrastive samples.
\end{itemize}

\vspace{-0.2cm}
\subsubsection{Authentic human enactment and quality control}
During the recording phase, 52 speakers enacted the structured scripts in accordance with the designated scenarios. The enactment intentionally preserved human behavioral variations, naturally embedding speech-rate shifts, pauses, sighs, and hesitations to closely approximate daily conversations. The foundational test set was finalized following baseline denoising and rigorous manual inspection to exclude samples with severe noise or mispronunciations.

\vspace{-0.2cm}
\subsubsection{Multiple-choice reformulation and distractor construction}
To mitigate the subjectivity of LLM-as-a-Judge evaluations in assessing comprehension and reasoning capabilities, the assessment of contextual and cross-modal understanding is transformed into objective MCQs. The audio for all multiple-choice options is synthesized using Qwen3-TTS~\cite{qwen3tts}, which employs the original speaker's voice as the reference audio to maintain acoustic consistency. These synthesized options are directly appended to the final turn of the human-recorded dialogue. Task~3 remains an open-ended generation task and therefore requires no options. For each sample in the objective tasks (Tasks~1, 2, and 4), three adversarial distractors are designed:
\begin{itemize}
    \item \textbf{Task~1 (Trajectory):} Distractors are formulated by flipping emotional polarity, scrambling temporal order, or substituting specific sequence segments. They appear locally plausible but remain globally inconsistent with the true emotional trajectory.
    \item \textbf{Task~2 (Reasoning):} An auxiliary LLM generates options attached to secondary facts explicitly mentioned in the text, yet detached from the genuine emotional trigger. This forces the evaluated ALM to distinguish mere factual statements from actual emotional causes.
    \item \textbf{Task~4 (Conflict):} A ``literal interpretation'' option based entirely on text semantics serves as a strong distractor. This option specifically exposes whether a model improperly prioritizes textual meaning over authentic acoustic emotion.
\end{itemize}

\vspace{-0.2cm}
\subsection{Evaluation methodology}
Tasks~1, 2, and 4 are evaluated automatically using standard classification accuracy. The generative task (Task~3) is assessed on a 1--5 scale across three defined dimensions. Dimension 1 (D1, Textual Empathy \& Insight) evaluates the depth of textual empathy, assessing whether the response actively validates and explores the user's emotional state rather than merely mirroring it. Dimension 2 (D2, Vocal Empathy \& Congruence) assesses paralinguistic appropriateness, evaluating whether the synthesized speech exhibits natural tonal shifts and warmth that correspond to the emotional context. Dimension 3 (D3, Audio Quality \& Naturalness) measures the technical clarity, fluency, and human-like naturalness of the synthesized audio. Detailed evaluation rubrics and prompts are available in the open-source repository\footnote{\urlstyle{same}\url{https://github.com/ASLP-lab/Hum-Dial}}. For scoring, D1 relies on advanced text LLMs (Qwen3-Omni and Gemini-2.5-Pro). Since current automated tools remain unreliable for quantifying complex vocal expressions, D2 and D3 are assessed by human judges. Ten experienced listeners evaluated both the Chinese and English subsets under a blind rating protocol.

\vspace{-0.2cm}
\section{Experiments}

\vspace{-0.2cm}
\subsection{Experimental setup}
We evaluate eight ALMs in two categories. The open-source group includes Freeze-Omni~\cite{freeze-omni}, GLM-4-Voice~\cite{glm4}, Kimi-Audio~\cite{kimi}, Step-Audio-2-mini~\cite{step-audio2}, and Qwen2.5-Omni~\cite{Qwen2.5-Omni}. The closed-source group consists of Doubao-realtime, GPT-4o-audio~\cite{gpt4o}, and Gemini-2.5-flash~\cite{gemini2.5}. These models represent the current state of the art in ALM performance.

\begin{table}[t]
  \caption{Accuracy (\%) on Tasks~1 and~2. \textbf{Bold}: best in column; \underline{underline}: second best.}
  \label{tab:objective_results}
  \centering
  \footnotesize
  \setlength{\tabcolsep}{4pt}
  \begin{tabular}{lcccccc}
    \toprule
    \multirow{2}{*}{\textbf{Model}}
      & \multicolumn{3}{c}{\textbf{Task 1 (Trajectory)}}
      & \multicolumn{3}{c}{\textbf{Task 2 (Reasoning)}} \\
    \cmidrule(lr){2-4} \cmidrule(lr){5-7}
      & CN & EN & Avg & CN & EN & Avg \\
    \midrule
    \multicolumn{7}{l}{\textit{Open-Source}} \\
    Freeze-Omni       & 16.00 & 6.00  & 11.00             & 23.88 & 24.16 & 24.02 \\
    GLM-4-Voice       & 42.67 & 18.00 & 30.34             & 58.96 & 52.35 & 55.66 \\
    Kimi-Audio        & 70.00 & 62.67 & 66.34             & 71.64 & 65.77 & 68.71 \\
    Step-Audio-2-mini & 34.67 & 34.67 & 34.67             & 67.16 & 76.51 & 71.84 \\
    Qwen2.5-Omni      & 66.67 & 80.67 & 73.67 & 76.09 & 77.85 & \underline{76.99} \\
    \midrule
    \multicolumn{7}{l}{\textit{Closed-Source}} \\
    Doubao-realtime   & 76.67 & 60.00 & 68.33             & 52.24 & 53.69 & 52.97 \\
    GPT-4o-audio      & \underline{79.33} & \underline{86.67} & \underline{83.00}             & \underline{72.39} & \underline{80.54} & 76.47 \\
    Gemini-2.5-flash  & \textbf{84.67} & \textbf{91.33} & \textbf{88.00} & \textbf{76.12} & \textbf{83.22} & \textbf{79.67} \\
    \bottomrule
  \end{tabular}
\end{table}

\vspace{-0.2cm}
\subsection{Results and analysis}

\textbf{Multi-Turn Emotional Tracking and Reasoning} As shown in Table~\ref{tab:objective_results}, closed-source models (e.g., Gemini-2.5-flash) achieve the highest average accuracies in emotional trajectory tracking (88.00\%) and implicit causal reasoning (79.67\%) for Tasks~1 and 2. Open-source models display high variance, with only Qwen2.5-Omni performing competitively. Notably, the two model classes exhibit opposite trends: closed-source models excel at trajectory tracking, demonstrating robust sequential emotion memory, whereas open-source models perform better at localized causal extraction but struggle with multi-step sequences. This disparity likely arises from differences in long-term context processing capacity.

\begin{table}[t]
  \caption{Scores on Task~3. D1: text empathy, scored by LLM judges (Qwen3-Omni / Gemini2.5-Pro); D2/D3: acoustic empathy / naturalness. \textbf{Bold}: best in column; \underline{underline}: second best; \textit{italics}: anomalous low.}
  \label{tab:generative_results}
  \centering
  \footnotesize
  \setlength{\tabcolsep}{3pt}
  \begin{tabular}{lcccc}
    \toprule
    \multirow{2}{*}{\textbf{Model}}
      & \multicolumn{2}{c}{\textbf{D1 (Q / G)}}
      & \multicolumn{2}{c}{\textbf{D2 / D3 (Human)}} \\
    \cmidrule(lr){2-3} \cmidrule(lr){4-5}
      & CN & EN & CN & EN \\
    \midrule
    \multicolumn{5}{l}{\textit{Open-Source}} \\
    Freeze-Omni       & \underline{3.69} / 2.69 & 3.25 / 2.52 & 3.29 / 3.42 & 3.37 / 3.45 \\
    GLM-4-Voice       & 3.62 / 3.38 & 2.77 / 2.42 & 3.27 / 3.31 & 3.37 / 3.41 \\
    Kimi-Audio        & 3.62 / 3.59 & \textit{1.17} / \textit{1.19} & 3.52 / 3.60 & 3.50 / 3.49 \\
    Qwen2.5-Omni      & 3.41 / 3.19 & 3.37 / 2.87 & \underline{3.55} / \underline{3.65} & 3.49 / \underline{3.59} \\
    Step-Audio-2-mini & 3.48 / 3.48 & 3.29 / 3.07 & 3.36 / 3.47 & 3.47 / 3.52 \\
    \midrule
    \multicolumn{5}{l}{\textit{Closed-Source}} \\
    Doubao-realtime   & 3.24 / 3.29 & 3.23 / 3.04 & \textbf{3.79} / \textbf{3.79} & \underline{3.58} / 3.57 \\
    GPT-4o-audio      & \textbf{3.77} / \textbf{3.79} & \textbf{3.77} / \textbf{4.10} & 3.23 / 3.26 & 3.42 / 3.49 \\
    Gemini-2.5-flash  & 3.68 / \underline{3.73} & \underline{3.64} / \underline{3.93} & 3.26 / 3.39 & \textbf{3.60} / \textbf{3.60} \\
    \bottomrule
  \end{tabular}
\end{table}

\noindent\textbf{Empathetic Response Generation} Table~\ref{tab:generative_results} presents three findings for Task~3. First, dual LLM evaluation of text empathy (D1) exposes significant scoring variance (e.g., a 1.0-point difference for Freeze-Omni), indicating subjectivity in current LLM-as-a-judge approaches. Second, despite systemic fluctuations in D1 scores, incorporating human-evaluated acoustic empathy (D2) demonstrates a consistent structural decoupling between textual and acoustic empathy. For example, Doubao-realtime achieves the highest acoustic empathy (D2: 3.79) despite moderate text empathy (D1: 3.24/3.29). This indicates ALMs can synthesize expressive speech independently of semantic depth. Finally, Kimi-Audio exhibits anomalously low English D1 scores (1.17/1.19) due to severe cross-lingual interference, resulting in frequent Chinese responses to English prompts.

\begin{table}[t]
  \caption{Accuracy (\%) on Task~4. ``Conflict'' denotes samples where acoustic emotion contradicts text sentiment; ``Nonconf.'' denotes consistent samples. \textbf{Bold}: best in column; \underline{underline}: second best.}
  \label{tab:conflict_results}
  \centering
  \footnotesize
  \setlength{\tabcolsep}{4pt}
  \begin{tabular}{lcccc}
    \toprule
    \multirow{2}{*}{\textbf{Model}}
      & \multicolumn{2}{c}{\textbf{CN}}
      & \multicolumn{2}{c}{\textbf{EN}} \\
    \cmidrule(lr){2-3} \cmidrule(lr){4-5}
      & Conflict & Nonconf. & Conflict & Nonconf. \\
    \midrule
    \multicolumn{5}{l}{\textit{Open-Source}} \\
    Freeze-Omni       & 18.00 & 50.00 & 22.00 & 60.00 \\
    GLM-4-Voice       & 30.00 & 82.00 & 20.00 & 54.00 \\
    Kimi-Audio        & 48.00 & 80.00 & 38.00 & 80.00 \\
    Step-Audio-2-mini & 24.00 & \underline{82.00} & 26.00 & \underline{82.00} \\
    Qwen2.5-Omni      & 22.00 & \textbf{88.00} & 32.00 & \textbf{86.00} \\
    \midrule
    \multicolumn{5}{l}{\textit{Closed-Source}} \\
    Doubao-realtime   & \textbf{54.00} & 74.00 & 40.00 & 64.00 \\
    GPT-4o-audio      & 34.00 & 70.00 & \underline{46.00} & 68.00 \\
    Gemini-2.5-flash  & \underline{50.00} & 80.00 & \textbf{62.00} & 70.00 \\
    \bottomrule
  \end{tabular}
  \vspace{-0.2cm}
\end{table}

\noindent\textbf{Acoustic-Semantic Conflict Recognition} Task~4 evaluates robustness when acoustic emotions contradict text semantics. Table~\ref{tab:conflict_results} shows that all ALMs degrade significantly on conflict samples. Notably, models excelling on consistent samples suffer severe declines; for instance, Qwen2.5-Omni accuracy drops from 88.00\% to 22.00\% on Chinese conflict samples. This demonstrates a prevalent ``text-dominance bias,'' where models under cross-modal interference over-rely on literal text and select text-derived distractors instead of prioritizing authentic acoustic emotion. Even robust closed-source models like Gemini-2.5-flash exhibit this vulnerability. Consequently, current audio-native architectures retain the ``transcribe-then-understand'' bias of cascaded systems, establishing cross-modal conflict resolution as a primary challenge.

\vspace{-0.2cm}
\section{Discussion and Conclusion}

This paper introduces HumDial-EIBench, an objective evaluation framework utilizing authentic human-recorded dialogues to assess the emotional intelligence of ALMs. By reformulating open-ended scenarios into multiple-choice tasks, the benchmark successfully isolates multi-turn emotional memory and reasoning capacities from superficial generation proficiency for independent diagnosis. System-level evaluations reveal critical structural deficiencies in current architectures, notably a widespread ``text-dominance bias'' under cross-modal conflicts and a ``structural decoupling'' between textual and acoustic empathy. These findings indicate that existing models still primarily process audio as a transcription proxy rather than treating it as an independent emotional modality of equal importance to text, underscoring the urgent necessity of explicit cross-modal consistency training to achieve genuine multimodal alignment.

Nevertheless, this study exhibits certain limitations. First, regarding the text empathy evaluation of generated responses (Task~3, D1), the high variance among different LLM judges indicates that automatically and objectively quantifying empathy depth remains an unsolved open problem. Second, the acoustic-semantic conflict evaluation (Task~4) currently focuses on single-turn utterances, whereas real-world sarcasm or implicit emotions are often intertwined within complex multi-turn interactions. Future work will aim to expand multi-turn conflict scenarios and explore more stable, reliable automatic evaluation metrics for multimodal empathy.

\section{Generative AI Use Disclosure}
Generative AI models, including Qwen3-TTS, Qwen2.5-Omni and Gemini 2.5 Pro, were used for data generation, and response evaluation. The authors are fully responsible and accountable for the final content of this paper. All authors agree with the submission of this paper.

\bibliographystyle{IEEEtran}
\bibliography{mybib}

\end{document}